\documentclass[twocolumn,prb,floatfix,showpacs,superscriptaddress]{revtex4-1}
\usepackage{amsmath,amssymb,graphicx,epstopdf,eps fig,subfigure}

\begin{document}

\author{Magnus O.\ Borgh}
\affiliation{School of Mathematics, University of Southampton,
  Southampton, SO17 1BJ, United Kingdom}
\author{Guido Franchetti}
\affiliation{Department of Applied Mathematics and Theoretical
  Physics, University of Cambridge, Cambridge, CB3 0WA, United
  Kingdom}
\author{Jonathan Keeling}
\affiliation{SUPA, School of Physics and Astronomy, University of St.\
  Andrews, KY16 9SS, United Kingdom} 
\author{Natalia G.\ Berloff}
\affiliation{Department of Applied Mathematics and Theoretical
  Physics, University of Cambridge, Cambridge, CB3 0WA, United Kingdom}

\title{Robustness and observability of rotating vortex-lattices in an
  exciton-polariton condensate}

\begin{abstract}
  Exciton-polariton condensates display a variety of intriguing
  pattern-forming behaviors, particularly when confined in
  potential traps.  It has previously been predicted that
  triangular lattices of vortices of the same sign will form 
  spontaneously as the result of surface instabilities in a
  harmonic trap.  
  However, natural disorder, deviation of the external
  potential from circular symmetry, or higher-order terms modifying
  the dynamical equations may all have detrimental effects and
  destabilize the circular trajectories of vortices.  
  Here we
  address these issues, by characterizing the robustness of the
  vortex lattice against disorder and deformations of the
  trapping potential.  
  Since most experiments use time integrated
  measurements it would be hard  to observe directly the rotating
  vortex lattices or distinguish them from vortex-free states.
  We suggest how these difficulties can be overcome and  present an
  experimentally viable 
  interference-imaging scheme that would allow the detection of
  rotating vortex lattices.
\end{abstract}

\maketitle

\section{Introduction}
\label{sec:introduction}

The experimental realization of Bose-Einstein condensates of
quasiparticles in solid-state
systems
has provided a new example of macroscopically coherent quantum states
of matter.  Solid-state condensates exhibit striking fundamental
differences from traditional quantum fluids such as atomic Bose gases
and superfluid liquid helium.  Much recent
interest\cite{kasprzak06:nature,balili2007:science,lagoudakis08,utsunomiya08,amo09:bullet,amo09:sf,lagoudakis09:science,wertz2010:nphys,sanvitto2010:nphys,nardin2010:prb,roumpos10:nphys,tosi11:prl,lagoudakis11:prl,manni2011:prl,nardin2011:nphys,grosso2011:prl,Tosi:2012ika}
has been devoted to exciton-polaritons, which exist as normal 
modes of strongly coupled excitons and photons in semiconductor
microcavities.  Due to the short lifetime of the quasiparticles, these
condensates exist in a dynamic balance between pumping and decay
rather than in true thermal equilibrium. As such, these systems are
capable of spatial and temporal pattern formation similar to that
exhibited by lasers.\cite{Arecchi:1999tl,staliunas03}

Hallmark features of superfluidity such as quantized
vortices\cite{lagoudakis08,lagoudakis09:science,nardin2010:prb,sanvitto2010:nphys,roumpos10:nphys,tosi11:prl,lagoudakis11:prl,nardin2011:nphys,grosso2011:prl}
and persistent currents\cite{sanvitto2010:nphys} have been observed in
polariton condensates.  Spatial patterns and soliton dynamics have
been seen to arise spontaneously in both one\cite{wertz2010:nphys} and
two\cite{manni2011:prl,Tosi:2012ika} dimensions. Intrinsic disorder in
the material\cite{Savona2007} or deliberately designed
potentials\cite{balili2007:science} induce currents that may further
facilitate symmetry breaking in the polariton system leading to novel
patterns. In particular, vortices have been seen to nucleate and get
trapped because of the nonuniformities of the potential
landscape.\cite{nardin2010:prb,nardin2011:nphys}

Most of the models that have been successfully and extensively used to
describe pattern formation in polariton condensates are based on the
complex Ginzburg-Landau equation (cGLE), which is also used in the
theory of lasers and thus illustrates the similarities between the two systems.
Such models have been applied to vortex nucleation and
motion,\cite{pigeon2011:prb,fraser2009:njp,flayac2012:prb,marchetti2010:prl,grosso2011:prl}
spatial density modulations,\cite{manni2011:prl,wouters2010:prb}
solitons\cite{grosso2011:prl} and vortex
lattices.\cite{keeling08:gpe,liew2008:prl,borgh10:prb,gorbach2010:prl}

Various different modifications of this model have been considered,
such as including separate dynamics of the reservoir, or modeling
processes of thermal relaxation.  Under the assumption of fast
relaxation of the reservoir and small condensate densities
these modifications all simplify and the model can be written as:
\begin{equation}
  \label{eq:cgpe}
  \begin{split}
    i \frac{\partial\psi}{\partial t} = \frac{1}{2} \Big[& - \nabla ^2 +
      V(x,y) +| \psi | ^2 \\ 
      &+ i \Big( \alpha (x,y)  - \sigma |\psi| ^2  \Big)  \Big ] \psi  
  \end{split}
\end{equation} 
where $V(x,y)$ describes a trapping potential, $\alpha(x,y)$ is
a (spatially varying) pump rate, and $\sigma$ is a nonlinearity
which causes pumping to reduce as density increases---some such
nonlinear process is clearly necessary for the system to be stable,
and the form included is the simplest such nonlinearity. Linear losses are included in $\alpha$.
Anticipating that we will consider a harmonically confined
system, Eq.~(\ref{eq:cgpe}) is stated in oscillator units, measuring
energy in units of 
$\hbar\omega$, length in units of the oscillator length
$l=\sqrt{\hbar/m\omega}$, and time in units of $\omega^{-1}$, where
$\omega$ is the oscillator frequency of the trapping potential, and
$m$ is the effective mass 
of the polariton. The values assumed in the numerical simulations are
discussed at the start of Sec.~\ref{sec:prop-rotat-state}.

The polarization degree of freedom of the polaritons provides
additional possibilities in pattern and defect formation, such as
allowing for spatially separate vortices of left and right
polarization.\cite{lagoudakis09:science} Further possibilities for
pattern formation arise in the presence of a magnetic field that
favors circular polarization and therefore competes with the
interactions, which favor equal densities of the polarization
components, and also with any anisotropy resulting from strain fields
induced by mechanical stress. The cGLE can be modified to
include these effects,\cite{borgh10:prb} but in this paper we
will consider linear polarization only.

In the absence of pumping and decay terms, the steady states of
the cGLE are often well approximated by the stationary Thomas-Fermi
solution.  One rather surprising feature of the cGLE is that there
exist many cases where the presence of pumping and decay significantly
modifies the stable density profile.  A particularly surprising case
is that in a harmonic trap, the stationary state
is unstable, and a rotating lattice is formed as the result of
this instability.\cite{keeling08:gpe}  Spontaneously formed
vortices have been observed in polariton
condensates,\cite{lagoudakis08,lagoudakis09:science,roumpos10:nphys,lagoudakis11:prl,tosi11:prl}
but these have been isolated, stationary
vortices, pinned by the disorder potential in the sample. The
spontaneous formation of a regular, rotating vortex
lattice\cite{keeling08:gpe,borgh10:prb} has not been seen
experimentally.  The aim of this paper is therefore to study the
robustness of such lattices, and to discuss how one may overcome the
difficulties that might arise in detecting a rotating vortex lattice.

We consider the robustness of vortex lattices with respect both to
deformations of the circular harmonic trapping potential considered
previously,~\cite{keeling08:gpe,borgh10:prb} and 
also to modifications of the cGLE, such as energy relaxation.  We find
that there is definite 
sensitivity to the trapping potential---both elliptical distortion
as well as background disorder can destroy the regular vortex lattice
and replace it with a chaotic regime, and we present numerical results
for the critical distortion and disorder that destroy the
lattice. This fragility with respect to the geometry of the trapping
potential contrasts with the relative robustness of the vortex 
lattice to other terms that might be present in the cGLE, such as
phenomenological approaches to the inclusion of relaxation processes.

Detection of moving vortices in a polariton condensate is non-trivial.
Polariton condensates can be imaged directly using the photons
escaping from the semiconductor microcavity.  Each image requires that
light be gathered over a non-negligible period of time, and therefore
any moving feature will tend to smear out. Hence rotating
configurations would not be directly visible in time integrated
images.  We show that these problems can be overcome by using a
defocused homodyne imaging scheme, allowing stationary repeatable
images to be observed, which fully characterize the vortex lattice.

The remainder of this paper is organized as follows.
Section~\ref{sec:inst-non-rotat} discusses the instability of the
stationary state, showing how this instability can be found
analytically for weak pumping and decay.
Section~\ref{sec:prop-rotat-state} then considers the vortex
lattices that result from this linear instability, presenting
results for elliptical trapping potentials, and traps with
additional disorder present.  Section~\ref{sec:detect-rotat-vort}
presents schemes for detecting rotating vortex lattices, and
section~\ref{sec:conclusions} summarizes these results.

\section{Instability of the non-rotating state }
\label{sec:inst-non-rotat}

As commented above, the presence of pumping and decay destabilizes the
stationary Thomas-Fermi potential in a circular trap, which in turn
leads to the appearance of the vortex lattice.\cite{keeling08:gpe}
In this section we discuss how, for weak pumping and decay, this
instability can be found analytically, providing some insight into how
and why the stationary state becomes unstable.  This section provides
an outline and the main results of the analytical stability
calculation; further technical details are given in
Appendix~\ref{sec:line-stab-analys}.

\subsection{Linear stability analysis}
\label{sec:line-stab-analys-1}

In order to find the
leading order effects of pumping and decay, one may
linearize both in fluctuations and in the effects of pumping
and decay, i.e., one considers first the density profile and
fluctuations about this in the absence of pumping and decay, and then
determines how these are modified by non-zero $\alpha$ and $\sigma$.

Using $\psi =\sqrt{\rho} e^{i \phi}$ and $V(x,y)=x^2+y^2\equiv r^2$, one may rewrite
Eq.~(\ref{eq:cgpe}) as a pair of real equations:
\begin{equation}
  \label{eq:madellung}
  \begin{aligned}    
  \partial_t \rho + \nabla\cdot(\rho \vec{v})
  = (\alpha - \sigma \rho) \rho
  \\
  2 \partial_t \vec{v} + \nabla ( \rho + r^2 + |\vec{v}|^2) = 0,
  \end{aligned}
\end{equation}
where $\vec{v} = \nabla \phi$.  In these equations, an approximation
has already been made in neglecting the quantum pressure terms; this
approximation
is appropriate for sufficiently smooth density profiles (i.e., for
clouds much larger than the healing length $\xi_\rho=1/\sqrt{\rho}$).  
By substituting $\rho \to \rho + h(t), \vec{v} \to \vec{v} + \vec{u}(t)$
into Eqs.~(\ref{eq:madellung}), and linearizing in $\vec{u}, h$,
the fluctuations obey
\begin{equation}
  \label{eq:linearised}
  \begin{aligned}
    \partial_t h 
    &=
    -
    \nabla \cdot (\rho \vec{u} + h \vec{v})
    +
    \left( \alpha - 2\sigma \rho\right) h  
    \\
    \partial_t \vec{u} &= - \nabla
    \left( h/2 + \vec{v} \cdot \vec{u} \right).
  \end{aligned}
\end{equation}
In the absence of pumping and decay, the steady state is given by
$\rho = \rho_{\rm TF} \equiv (\mu - r^2)\Theta(\mu - r^2)$, where
$\Theta$ is the unit step function and  $\vec{v}=0$,  so the terms in
Eq.~(\ref{eq:linearised}) involving $\vec{v}$ vanish.  
In this case
Eq.~(\ref{eq:linearised}) can be reduced to the single
equation
\begin{equation}
  \label{eq:hypgeom}
    \frac{d^2 h}{d t^2}  = \frac{1}{2r} \frac{d}{dr} \left(
      r \rho \frac{dh}{dr} \right) + \frac{\rho}{2 r^2} 
    \frac{d^2}{d\theta^2} h,
\end{equation}
which has eigenfunctions $i \partial_t h_{n,s} = \omega_{n,s} h_{n,s}$
labeled by angular momentum $h_{n,s}(r,\theta) = e^{i s \theta}
h_{n,s}(r)$, where $h_{n,s}(r)$ are hypergeometric functions of the
radial variable $r$ 
(see Appendix~\ref{sec:exact-zeroth-order}).  These eigenfunctions have
frequencies
\begin{equation}
  \label{eq:eigenfrequencies}
  \omega_{n,s} = \sqrt{ s (1+2n) + 2n (n+1)}.
\end{equation}

\subsection{Perturbative shift with pumping and decay}
\label{sec:pert-shift-with}

We may now use the solutions~(\ref{eq:hypgeom}) and
(\ref{eq:eigenfrequencies}) as a basis on which to consider
perturbatively the effects of pumping and decay.  We will work here to
first order in $\alpha$ and $\sigma$, in order to see the qualitative
change introduced by pumping.  This first-order treatment will predict an
instability that already explains the behavior seen when all orders are
considered in the numerical results 
presented in Ref.~\onlinecite{keeling08:gpe} and in
Sec.~\ref{sec:prop-rotat-state}.

At first
order,\cite{keeling08:gpe} the steady state of
Eq.~(\ref{eq:madellung}) is given by $\rho = \rho_{\rm TF}$ and $\vec{v} =
-\sigma \rho_{\rm TF} \vec{r}/6$. 
Integrating the first of Eq.~(\ref{eq:madellung}) over the entire
space gives $\mu = 3 \alpha/2 
\sigma$. To find the change to the frequencies of the fluctuations,
it is helpful to consider Eq.~(\ref{eq:linearised}) in the form
\begin{equation}
  \label{eq:perturbation}
  (\omega_{0i} + \delta \omega_i) 
  (\psi^R_{0i} + \delta \psi^R_i)
  =
  i(\mathcal{L} + \alpha \mathcal{\delta{L}})
  (\psi^R_{0i} + \delta \psi^R_i),
\end{equation}
where $\psi^R_i$ represent the right eigenfunctions of the matrix of
differential operators $i \mathcal{L}$.  Note that in this
identification, the terms involving $\vec{v}$ in
Eq.~(\ref{eq:linearised}) should be considered part of $\delta
\mathcal{L}$.  The change to the eigenfrequency induced by pumping and
decay is thus
\begin{equation}
  \label{eq:change}
  \delta \omega_i = i\alpha 
  \left.
    \left[
    \int d^2r
    \psi^L_{0i} \delta\mathcal{L} \psi^R_{0i}
    \right]
  \middle/
  \left[
    \int d^2r
    \psi^L_{0i}  \psi^R_{0i}
    \right],
  \right.
\end{equation}
where $\psi^L_i$ indicate left eigenfunctions.  Finding left and right
eigenfunctions separately is necessary as the
operator $i\mathcal{L}$ is not self-adjoint; further details are
given in Appendix~\ref{sec:pert-theory-non}.

\subsubsection{Instability of the stationary solution}
\label{sec:inst-stat-solut}

Using the
hypergeometric solutions described above, and following the outlined procedure
(see Appendix~\ref{sec:eval-overl-hyperg} for details), one finds the shift
\begin{equation}
  \label{eq:im-shift}
  \delta \omega_i 
  = 
  \frac{i \alpha}{4}
  \left[\frac{s^2 -  s(1+2n) - 2n(n+1)}{\frac{1}{2}s^2 + s(1+2n)+2n(n+1) } \right].
\end{equation}
The crucial feature of this result is that for any $n$ there is always $s$ (for instance $s
\gg n$), such that  this expression is positive.  This implies that the
eigenfrequencies acquire a positive imaginary part
and the corresponding eigenmodes, which are those with large angular
momentum, will therefore grow, 
leading to vortex nucleation
(cf.\ atomic Bose-Einstein condensates where vortices are nucleated
when external 
stirring causes unstable eigenmodes to arise\cite{sinha2001:prl}).

\subsubsection{Including phenomenological relaxation}
\label{sec:incl-phen-relax}

The same procedure may be repeated includinging various modifications
to the cGLE.  One such modification that may  be easily incorporated is
an energy-relaxation term, causing time evolution toward lower energy
states, i.e., $\partial_t \psi \to (1+ i \eta) \partial_t
\psi$. The introduction of this term is common in literature on
atomic Bose-Einstein condensates, where it models the atomic
transfer between the 
condensate and the thermal cloud.  These effects can be accurately
modeled by a quantum-Boltzmann equation describing the population 
dynamics of the quantum states. At lowest order of
approximation this leads to the same
ansatz.\cite{Penckwitt:2002kda,Gardiner:1997jca}
Such terms have also
been used in modeling polariton
condensates.\cite{wouters10:superfluid,wouters10:relax,wouters2010:prb} 
As described in Appendix~\ref{sec:line-stab-gener}, including such an
effect at linear order changes the steady state chemical potential to
$\mu = 3 \alpha/(2 \sigma + 3 \eta)$, and the frequency shift becomes
\begin{equation}
  \label{eq:im-shift-eta}
  \delta \omega_i 
  = 
  \frac{i \alpha}{4}
  \left[\frac{\frac{2 \sigma}{2\sigma + 3 \eta}s^2 -  s(1+2n) - 2n(n+1)}{\frac{1}{2}s^2 + s(1+2n)+2n(n+1) } \right].
\end{equation}
Thus such relaxation, although driving the cGLE toward its low-energy
states, is not sufficient to kill the instability---for any $n$ and any
finite $\eta/\sigma$, there always exist modes (for instance, with $s \gg n
\sigma/\eta$) which will become unstable.

\section{Properties of the rotating state}
\label{sec:prop-rotat-state}

The above stability analysis assumed an infinite pump
spot, however the same instability still exists as long as the
pumping spot is larger than the condensate cloud size (given by
$r_\mathrm{TF} = \sqrt{\mu} = \sqrt{3 \alpha/2\sigma}$).  
As found previously in Ref.~\onlinecite{keeling08:gpe}, the 
instability leads to a rotating vortex lattice in a circular pump spot.
In this section 
we discuss the appearance of rotating states in cases where the
trapping and pumping deviate from the ideal circular harmonic case
considered in previous works.\cite{keeling08:gpe,borgh10:prb}  To
discuss such cases, we present results of numerical simulations.

In choosing values of the parameters $\alpha$ and $\sigma$ we follow the
numbers used in Refs.~\onlinecite{keeling08:gpe} and
\onlinecite{borgh10:prb}.  The parameter $\sigma$  may be directly
related to the blueshift of the condensate when pumping at a given
multiple of the threshold power. In this way $\sigma$ is estimated
in Ref.~\onlinecite{keeling08:gpe} as $\sigma=0.3$.  The
dimensionless gain $\alpha$ is given by the background linewidth
$\kappa$ and the pumping power $\gamma$ as
$\alpha=2(\gamma-\kappa)/\hbar\omega$. Estimating the polariton
lifetime at $\tau\simeq5$~ps gives a linear decay rate
$\kappa=\hbar/\tau=0.13$~meV. To facilitate comparison with
Refs.~\onlinecite{keeling08:gpe} and \onlinecite{borgh10:prb}, we
choose $\alpha=4.4$. In the stress trap constructed by Balili et
al,\cite{balili2006:apl,balili2007:science}
$\hbar\omega=0.066$~meV for the harmonic confinement under the
stressor, and the effective mass of the polariton is $m=7\times10^{-5}m_e$.
For this value of $\hbar \omega$ our chosen $\alpha$
roughly corresponds to pumping at twice the threshold.

\subsection{Non-circular geometry}
\label{sec:non-circ-geom2}

\begin{figure}[tbp]
  \begin{centering}
    \includegraphics[scale=1]{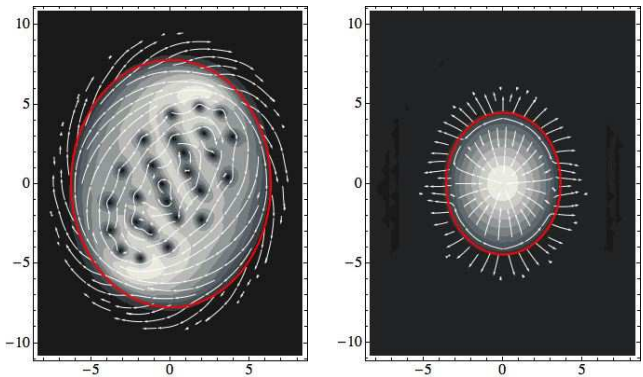}
    \caption{(color online) Density (grayscale) and superfluid streamlines
      (arrows) for different sizes of the pumping spot (red ellipse). 
      Left: ($\delta=0.9$, $R=7$) The surface instability leads to a
      spontaneously formed lattice with vortices 
      moving on elliptical trajectories.  
      Right: ($\delta=0.9$, $R=4$) The pumping spot is smaller than
      $r_\mathrm{TF}$ and the instability does not appear.}
    \label{fig:vortarray}
  \end{centering}
\end{figure}

Earlier considerations of spontaneously formed rotating vortex
lattices in harmonic potentials\cite{keeling08:gpe,borgh10:prb} have
assumed circular symmetry. Here we lift this assumption by considering
an elliptical geometry.  For simplicity in illustrating the
effects of elliptical geometry, we 
take both the trapping potential and the pumping spot in
Eq.~(\ref{eq:cgpe}) to be 
described by a parameter $\delta$, such that\footnote{The deformation
  $\delta$ is related to the eccentricity of the ellipse as 
  $\varepsilon =\sqrt{ 1-1/\delta^4}$}
\begin{equation} 
  V(x,y)=\frac{x^2}{\delta^2 }+ \delta^2 y^2.
\end{equation}
Using parameter values as detailed above, the pumping is taken to be
constant inside the ellipse 
$x^2/\delta^2 + \delta^2 y^2 = R^2$. The finite lifetime of the
polaritons leads to a linear loss outside the pumping
region. Since we are considering pumping at twice threshold,
$\gamma = 2 \kappa$, the value of $\alpha$ outside the pump spot is
the same as inside but with opposite sign.  As in the circular
case,\cite{keeling08:gpe} 
the surface instability will appear only if the pumping spot is
sufficiently large, as illustrated in Fig.~\ref{fig:vortarray} for
$\delta=0.9$.  For our subsequent analysis, we keep a fixed $R=7$.

\begin{figure}[tbp]
\begin{centering}
  \includegraphics[scale=1]{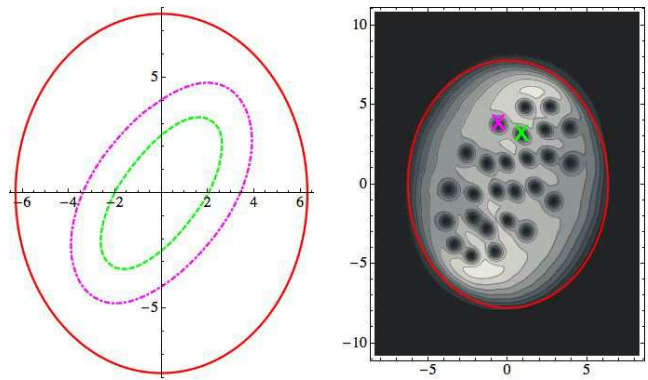}
  \caption{(color online) Left: the averaged orbits of two vortices
    (green, dashed and 
    magenta, dash-dotted lines, respectively).  The red, solid ellipse
    marks the pumping spot. Right: snapshot of the vortex lattice with
    the tracked vortices marked by green and magenta crosses.}
  \label{fig:ellorbits}
\end{centering}
\end{figure}

The movement of individual vortices in the lattice can be traced using
a minimum-finding algorithm.  
The orbit is obtained by averaging the trajectory over a number of
revolutions.  In the regular vortex lattice, the individual vortices
move on elliptic orbits as shown in Fig.~\ref{fig:ellorbits}. Because
the vortices have a definite sense of rotation, the motion is not
symmetric around the major axis, and the resulting elliptic vortex
lattice is tilted with respect to the trap.

\begin{figure}[tbp]
  \begin{center}
    \includegraphics[scale=1]{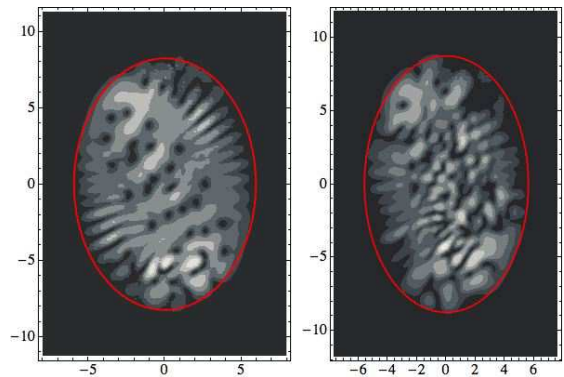}
    \caption{(color online) Left: $\delta=0.85$; the regular vortex
      lattice is broken up, 
      with some vortices still on regular elliptic orbits. 
      Right: $\delta=0.80$; as $\delta$  decreases, the rotating lattice gives
      way to a fragmented state with vortices continually entering
      and leaving the cloud.}
    \label{fig:breakdown}
  \end{center}
\end{figure}

In a sufficiently eccentric geometry, a steady array of vortices
fails to form.  We find a critical $\delta\sim0.85$ for which some
vortices move on elliptical orbits [Fig.~\ref{fig:breakdown}, left].
At $\delta=0.8$, the behavior is entirely chaotic, with vortices
continually entering and leaving the cloud [Fig.~\ref{fig:breakdown},
right], causing fragmentation of the condensate.

\begin{figure}[tbp]
  \begin{center}
    \includegraphics[scale=0.99]{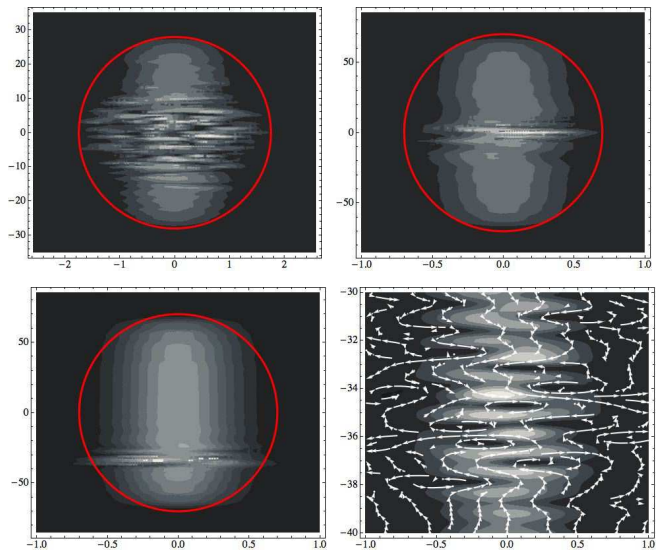}
    \caption{(color online) Top left: $\delta=0.25$. In a highly
      eccentric trap, the 
      density tries to establish a profile of approximately uniform
      width along the length of the trap.  Instability towards
      entrance of vortices leads to transverse oscillations.
      Note that the
      vertical axis has been rescaled so that the pumping spot
      appears circular in order to accommodate the extreme eccentricity.
      Top right: $\delta=0.1$. Extreme eccentricity suppresses fragmentation of
      the cloud.  Transverse oscillations develop at the midpoint.
      Bottom left: Spontaneously entering vortices cause an effective pressure
      gradient, pushing the region of perturbations towards one end of
      the trap.  The final position is shown. 
      Bottom right:  Vortices of both signs are present in the perturbed
      region. Arrows indicate circulation.} 
    \label{fig:eccentric}
  \end{center}
\end{figure}

In the limit of extreme eccentricity, the condensate tries to
establish a density profile of approximately uniform width in the
transverse direction, leading to a less fragmented condensate at
$\delta\lesssim 0.2$ [Fig.~\ref{fig:eccentric}].  The instability towards
entrance of vortices is still present where the spot is
widest: vortices entering and leaving the cloud cause transverse
oscillations.  As illustrated in Fig.~\ref{fig:eccentric}, vortices of
opposite signs may be present.  The distribution of vortices of
different sign may lead to a net effective pressure in the condensate
pushing the perturbation along the length of the condensate until the
pressure gradient disappears.  This effect is illustrated in
Fig.~\ref{fig:eccentric} (bottom), showing the final position of the region of
perturbations.  Note that a true stationary state is not obtained:
oscillations are still present.

We see that the formation of a regular vortex lattice is sensitive to
the geometry of the system, in contrast to the robustness against
refinements of the pumping model discussed in
Sec.~\ref{sec:incl-phen-relax} and Sec.~\ref{sec:other-effects}. While
a rotating vortex lattice will still form under small deviations from
circular geometry, moderate eccentricity will prevent the lattice from
forming.

\subsection{Effects of disorder on the rotating state}
\label{sec:disorder}

In the samples used in experiments disorder is always
present.\cite{Savona2007}  Here we take this into account by including
a disorder term in $V(x,y)$. This is modeled as a Gaussian random
field with real-space correlation function
\begin{equation}
  \label{eq:corr}
  \langle V_{\mathrm{dis}}(\vec{r})V_{\mathrm{dis}}(\vec{r}^{\,\prime})\rangle = 
  V_0^2 \exp\left(-\frac{(\vec{r}-\vec{r}^{\,\prime})^2}{2\xi^2}\right),
\end{equation}
where the amplitude $V_0$ is a measure of the strength of the disorder,
and $\xi$ is the correlation length. (Such a potential can be
constructed by sampling the Fourier components of the disorder potential from a
Gaussian distribution with a $k$-dependent variance; see
Appendix~\ref{sec:Vdis}.) 

\begin{figure}[tbp]
  \begin{center}
    \includegraphics[scale=0.8]{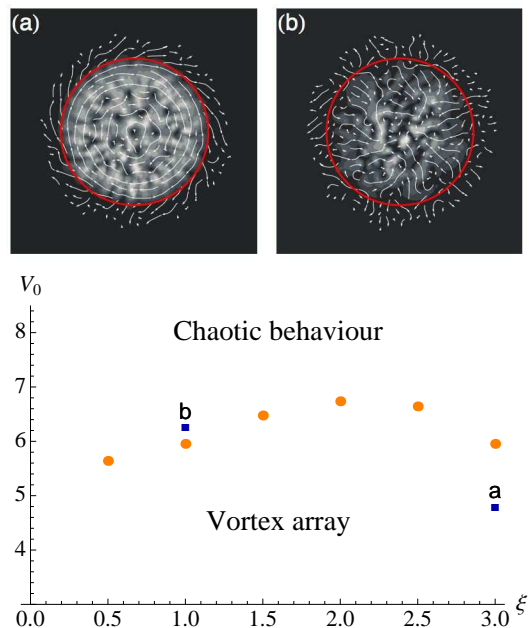}
    \caption{(color online) The rotating vortex lattice forms also in
      the presence 
      of weak disorder (a), but is destroyed by a disorder stronger
      than $V_{\rm crit}$, leading to a chaotic state (b). 
      Bottom: the boundary between vortex-lattice and chaotic regimes in
      the $(V_0,\xi)$ parameter space. The choices for $(V_0,\xi)$
      leading to states (a) and (b) are indicated in the diagram. 
      The vortex lattice is strongly 
      sensitive to the amount of disorder, whereas the correlation
      length is of only marginal importance.}
    \label{fig:disorder}
  \end{center}
\end{figure}

Varying $V_0$ and $\xi$ we find that the formation of a vortex
lattice is sensitive to the amount of disorder, with the lattice being
destroyed for $V_0 > V_\mathrm{crit} \simeq 6.5$.  
When the disorder strength exceeds $V_\mathrm{crit}$ the
behavior is chaotic and no stationary state forms.
However, the critical amount of
disorder is only weakly dependent on the disorder correlation length
$\xi$, as illustrated by the nearly horizontal border in the
$(V_0,\xi)$-diagram of Fig.~\ref{fig:disorder}. For the stress
trap of Ref.~\onlinecite{balili2007:science} our estimate 
translates to a critical disorder $V_\mathrm{crit}\simeq0.4$~meV,
subject to uncertainties in estimates of experimental quantities as
well as effects not accounted for in our model, such as 
relaxation mechanisms.

It is worth noting that $V_\mathrm{crit}>\hbar\omega$ may give
a false impression of a very large critical disorder strength. The relevant
energy scale is given by effective
depth of the harmonic potential over the extent of the cloud, and
this, in turn, is
determined by the chemical potential $\mu\gg\hbar\omega$.

\subsection{Effects of other modifications to the cGLE model}
\label{sec:other-effects}

There are a number of other possible modifications to the simple cGLE
model~(\ref{eq:cgpe}) that could potentially destabilize a vortex lattice or
prevent its formation: presence of repulsive interactions between
condensate particles and reservoir particles, presence of higher-order
nonlinear terms, or a finite gain linewidth leading to the
appearance of superdiffusion in the governing equations. In this
section we comment briefly on possible consequences of considering
these effects in the cGLE model.

\textit{Repulsive interactions with an excitonic reservoir.} Repulsive
interactions between reservoir polaritons and the condensate
have been predicted theoretically\cite{wouters07:bec} and detected
experimentally.\cite{Christmann:2012wc}  
For our geometry, such
interactions may be modeled by adding a Gaussian potential to the
center of the trap potential $V(x,y)$. The presence of this additional
potential term does not prevent lattice formation, but increases the
probability of having a central, stationary vortex, as the reservoir
provides a pinning site.  The presence of Gaussian reservoir repulsion may
further change the structure of the lattice: for example a 
sharp and narrow Gaussian favors a square, rather than triangular, lattice.

\textit{Higher order nonlinearities.} One may consider higher-order
corrections to polariton interactions in the form of a quintic
nonlinearity in the cGLE. In some cases, inclusion of quintic terms
in cGLE models is known to change the stability of the
solutions.\cite{aranson02}
We have checked that the
introduction of such terms in our model does not have any pronounced
effect on vortex lattice formation.

\textit{Superdiffusion}. Lasers emit particular transverse modes that
depend on the detuning between the longitudinal modes of the reservoir
(controlled by the resonator length) and the frequency at
which gain is maximum.\cite{Arecchi:1999tl} If
such gain selection is relevant to the polariton condensates the
right-hand side of Eq.~(\ref{eq:cgpe}) acquires a term
$i\nu(\nabla^2 + \Delta)^2\psi$, where $\Delta$ is the
detuning of the gain above the lowest transverse mode.
 Such mechanisms have not been extensively discussed in the
context of polariton condensation, but the essence appears for example
in Ref.~\onlinecite{doan06}. Our numerical simulations show that
although the lattice 
survives for small values of $\nu$ the rings of vortices are
shifted to the boundary of the condensate. The lattice disappears for
$\nu > 0.1$.

\section{Detecting rotating vortex lattices}
\label{sec:detect-rotat-vort}

We now discuss how time-integrated imaging techniques can be used to
observe a non-stationary vortex lattice in experiments. Images are
produced by gathering light escaping from the condensate over a period
of time long enough that the movement of the vortices 
is non-negligible.  We will construct a way of extracting an
image that remains stationary during the measurement despite this
movement.  In practice, imaging is done by combining 
images from several realizations of the experiment.  It is
therefore necessary that the same image is produced in each
realization.  In the following, we discuss
the relative merits of heterodyne and homodyne imaging techniques, 
before combining the desirable features of both into an experimentally
viable scheme described in Sec.~\ref{sec:defocused}.

\begin{figure}[tb]
  \centering
  \includegraphics[width=0.48\textwidth]{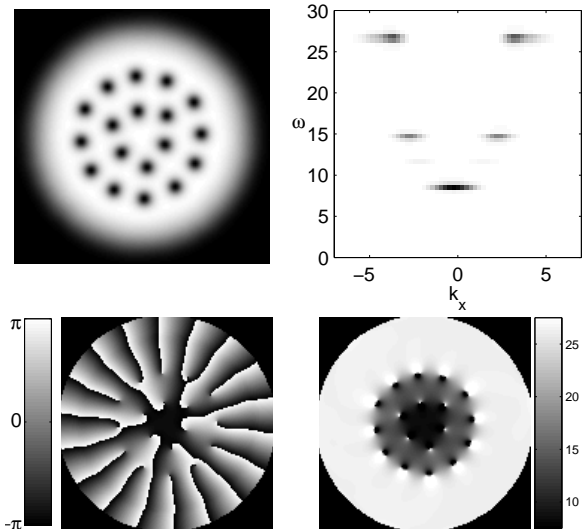}
  \caption{Example of a vortex-lattice solution to
    Eq.~(\ref{eq:cgpe}).  
    Top left: density, showing the vortex lattice.
    Top right: the corresponding spectrum.  Note the three 
    bands corresponding to the regions of the condensate separated by the 
    rings of vortices.
    Bottom: complex phase of the wavefunction (left), and the time
    derivative of the phase (right). Note the separation into regions
    with approximately constant time derivative corresponding to the
    bands in the spectrum.}
  \label{fig:lattice-spectrum}
\end{figure}

In the imaging schemes, we exploit the fact that a rotating vortex
lattice generates sharp sidebands in the spectrum.\cite{borgh10:prb} A
representative example is shown in
Fig.~\ref{fig:lattice-spectrum}. Each sideband corresponds exactly
to one of the regions of the condensate separated by the rings of
vortices. By interfering each region with light with matching
frequency, the phase difference between light from the condensate and
the interference beam will be constant in time, creating a fixed
interference pattern, which can be integrated over time without
smearing. (See Appendix~\ref{sec:interference-images} for
  technical details.)

\subsection{Heterodyne imaging}
\label{sec:heterodyne}

We first
consider a heterodyne interference scheme, in which the image of
the condensate is interfered with a monochromatic reference beam
without spatial structure, $\psi_{\rm ref} = 
A_0\exp[-i(\theta_0+\omega_0 t)]$. 
If $\omega_0$ is chosen to match
one of the sidebands in the spectrum, 
the phase difference between reference beam and
condensate is independent of time in the region corresponding to the
chosen sideband. Thus regions of constructive
and destructive interference in this part of the image remain
stationary. 

\begin{figure}[tb]
  \centering
  \includegraphics[width=0.48\textwidth]{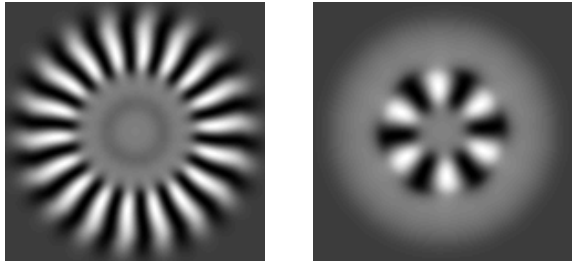}
  \caption{Time-integrated interference image using the heterodyne scheme and 
    the vortex-lattice from Fig.~\ref{fig:lattice-spectrum}.  
    Left: $\omega_0=26.5$.  Right: $\omega_0=14.5$.}
  \label{fig:heterodyne}
\end{figure}

Fig.~\ref{fig:heterodyne} shows interference images
obtained from the example state of Fig.~\ref{fig:lattice-spectrum}.
The vortex lattice appears in the image as interference fringes, the
number of fringes corresponding to the number of vortices inside the
stationary image.

This heterodyne scheme serves as a conceptually straightforward
illustration of the type of image we would like to obtain.  From an
experimental point of view, however, the scheme is impracticable.  A
fundamental problem is that the initial 
phase $\theta_0$ is arbitrary.  This means that while the interference
fringes remain stationary within one measurement,
there is no mechanism for ensuring that they appear in the same place
in any other realization (i.e., for locking the initial phase
difference).  This is crucial, as it means that while each image, in
principle, will be clear, averaging over several realizations will
cause the interference fringes to disappear.  In addition to this
fundamental problem, there may be practical difficulties in
establishing a sufficiently stable phase reference at exactly the
required frequency.

\subsection{Frequency-filtered homodyne imaging}
\label{sec:homodyne}

Since the heterodyne scheme has problems with the arbitrary phase
reference, we next consider using a homodyne scheme to avoid this. 
The simplest homodyne scheme that would reveal information about a
vortex lattice is one where an image of the condensate is filtered to
a narrow frequency band and then interfered with a retroreflected and
displaced copy of itself. This is a conceptually straightforward extension of
well-established experimental techniques for detecting vortices in
polariton condensates.\cite{lagoudakis08}

\begin{figure}[tb]
  \centering
  \includegraphics[width=0.48\textwidth]{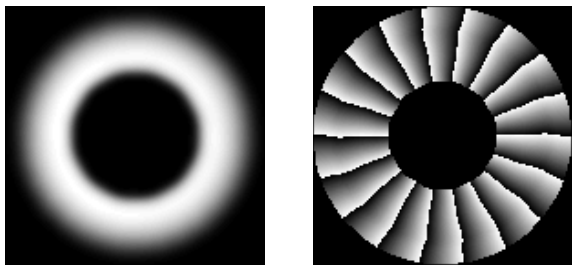}
  \caption{Snapshot of intensity (left) and phase (right) after 
    filtering $\psi$ 
    from Fig.~\ref{fig:lattice-spectrum} to the band 
    $26.25 \le \omega \le 26.75$}
  \label{fig:filter}
\end{figure}

Frequency filtering [Eq.~(\ref{eq:filter})] ensures that even after
displacement, each part of 
the image is interfered only with light of the same frequency, which
is necessary to create a stationary interference pattern. Density and
phase of the example lattice from 
Fig.~\ref{fig:lattice-spectrum} after filtering to
$26.25\leq\omega\leq26.75$ are shown in
Fig.~\ref{fig:filter}.

\begin{figure}[tb]
  \centering
  \includegraphics[width=0.48\textwidth]{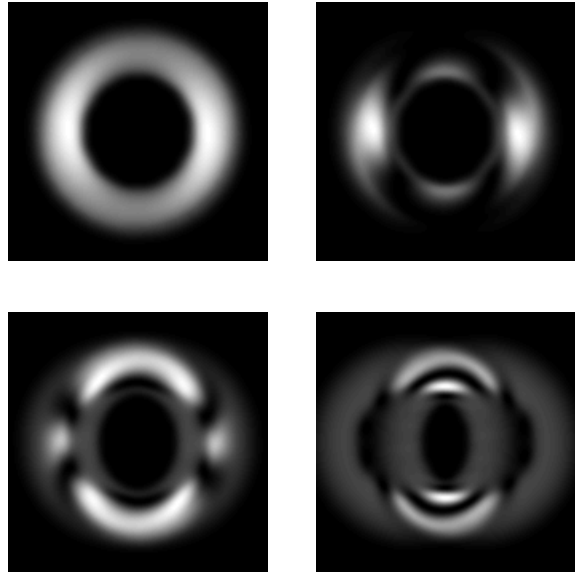}
  \caption{Frequency-filtered homodyne scheme applied to the solution in 
    Fig.~\ref{fig:lattice-spectrum} (the result of the filtering shown in 
    Fig~\ref{fig:filter}).  From top left to bottom right, the
    displacement [see Eq.~(\ref{eq:interference})]
    is $x_\mathrm{d}=0.2$, $x_\mathrm{d}=0.5$, $x_\mathrm{d}=1.0$, and
    $x_\mathrm{d}=2.0$. ($y_0=0$.)}        
  \label{fig:homodyne}
\end{figure}

The interference pattern created from Fig.~\ref{fig:filter} 
depends on the amount and direction of 
the displacement between the copies [Fig.~\ref{fig:homodyne}].  
The images can be understood by considering how overlapping areas of the
image in Fig.~\ref{fig:filter} and its retroreflected copy move in and out of
phase as displacement is varied. In Fig.~\ref{fig:homodyne}, this
leads to destructive interference for small displacement, and to a
complicated interference pattern as displacement is
increased.\footnote{In Fig.~\ref{fig:homodyne} an even number of
  vortices are enclosed by the filtered image. If an odd 
  number of vortices were enclosed, regions 
  of constructive and destructive interference would be interchanged.}

Since the image is produced by interfering light from the
condensate with itself, there is no arbitrary phase difference.
The resulting image is therefore identical between realizations, and
images from different realizations can be added without smearing.
However, interpretation of the images is difficult.  It is not
straightforward to infer the vortex lattice from the interference
image, let alone glean more detailed information about its
structure.

\subsection{Defocused homodyne imaging}
\label{sec:defocused}

We resolve the problems identified above by considering a defocused
homodyne scheme 
constructed to combine the desirable features of the
heterodyne and homodyne schemes. This will use a spatially uniform and
nearly monochromatic reference beam, but with a mechanism for locking
the arbitrary phase reference between realizations.

From the image of the condensate, light is 
taken from a reference point $\vec{r}_0$ and then defocused
and interfered with the full image.  Here we assume a
reference point without spatial extent, corresponding to one grid point in the
numerical simulation.  In practice, 
the reference point would be a fixed spot small enough that the spatial 
variation of the phase across it is negligible.  Optionally a
frequency filter [Eq.~(\ref{eq:filter})] may  
be applied as the first step of the scheme.

The scheme then amounts to forming the interference 
$I(\vec{r},t)=\left|\psi(\vec{r},t) + \psi(\vec{r}_0,t)\right|^2$.
Choosing $\vec{r}_0$ to be in a ring where
the time derivative of the phase is approximately constant (see
Fig.~\ref{fig:lattice-spectrum}) creates a
reference beam 
with the frequency of a side band in the spectrum.  Therefore, 
the interference image is qualitatively the same as that
produced by the heterodyne scheme,
but since the phase is now provided by the condensate itself, 
$\theta_0$ is no longer arbitrary. By construction, the interference
at $\vec{r}_0$ is always constructive, thus locking the
position of the interference fringes. 

\begin{figure}[tb]
  \centering
  \includegraphics[width=0.48\textwidth]{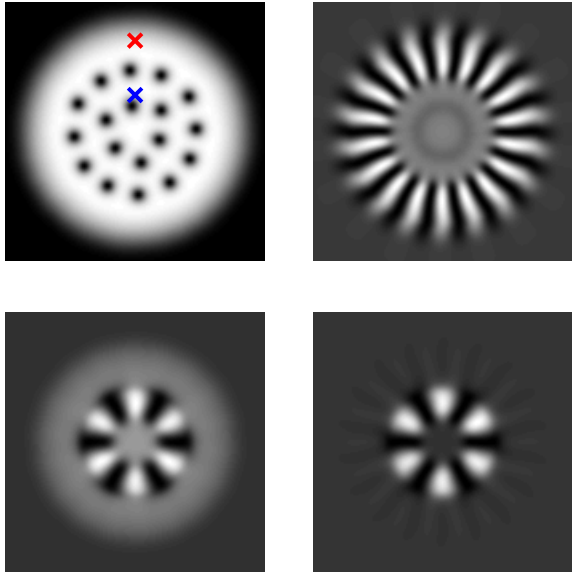}
  \caption{(color online) Top left: Snapshot of density with crosses
  showing the points  
  from which the reference is taken.  Top right: Interference with the 
  reference beam taken from the red cross, mapping out the interference 
  fringes of all 18 vortices enclosed by that part of the condensate. 
  Bottom left: Interference with light taken from the blue cross, giving 
  six stationary interference fringes.  Bottom right: Same as bottom left, 
  but with the image filtered to $14.15 \le \omega \le 14.75$ before the
  reference is extracted and allowed to interfere with the filtered image.}
  \label{fig:defocused}
\end{figure}

Fig.~\ref{fig:defocused} shows the defocused homodyne scheme applied
to Fig.~\ref{fig:lattice-spectrum}.  The reference point  
$\vec{r}_0$ is indicated on the snapshot of the density.  Selecting
$\vec{r}_0$ between the vortex rings yields an interference image with
six fringes corresponding  
to the inner ring of six vortices.  Selecting $\vec{r}_0$ near the edge
of the cloud reveals all 18 vortices.

The bottom panels of Fig.~\ref{fig:defocused} show the effect of 
frequency filtering prior to applying the defocused homodyne scheme. 
Without filtering, interference fringes are mapped out the same way as
in the heterodyne scheme.  However,  
filtering may yield a clearer picture with better contrast.

\begin{figure}[tb]
  \centering
  \includegraphics[width=0.48\textwidth]{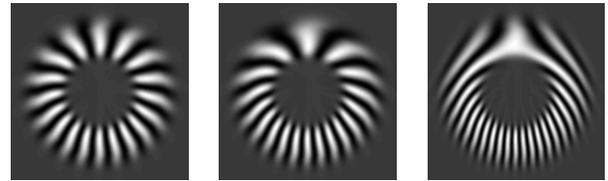}
  \caption{Interference with light taken from the point indicated by the red 
    cross in Fig.~\ref{fig:defocused} after filtering to a narrow frequency 
    band.  The panels show the resulting interference image for different 
    in-plane momenta in the $x$-direction.  From left to right
    $k_x=1.0, 2.0$ and $5.0$.} 
  \label{fig:defocused-finite-k}
\end{figure}

Slight misalignment (intentional\cite{lagoudakis08} or accidental) in
the interference optics may introduce a phase 
gradient in one of the pathways. In order to investigate whether the
proposed scheme is robust against such misalignment we introduce a  
finite in-plane momentum $\vec{k}$ in the reference beam, such that
$\psi_{\rm ref}(\vec{r},t) = \psi(\vec{r}_0,t)\exp(-i\vec{k}\cdot\vec{r})$.
This results in a distortion  
of the interference fringes [Fig.~\ref{fig:defocused-finite-k}]. The
distortion is small as long as $\vec{k}$ is not too large.

\begin{figure}[tbp]
  \centering
  \includegraphics[width=0.48\textwidth]{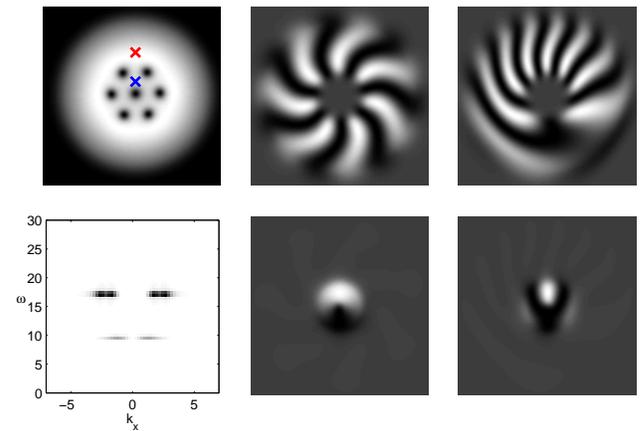}
  \caption{(color online) Top and bottom left: Density and the
  corresponding spectrum.   
  Crosses mark reference points for the filtered defocused homodyne scheme. 
  Remaining panels show interference images corresponding to light taken from 
  the red (top) and blue (bottom) crosses.  From left to right in both rows,
  in-plane momentum in the $x$-direction $k_x=0$, $k_x=1$ and $k_x=2$.}
  \label{fig:small-lattice}
\end{figure}

The distortion is more severe for smaller
vortex lattices, as demonstrated in Fig.~\ref{fig:small-lattice}.
Note that this lattice rotates in the opposite direction
compared with Fig.~\ref{fig:lattice-spectrum}, and that this is
reflected in the distortion of the fringes in the presence of the phase
gradient (cf.\ Fig.~\ref{fig:defocused-finite-k}).

\section{Conclusions}
\label{sec:conclusions}

We 
have shown how the instability of high-angular momentum modes leading
to spontaneous formation of vortex lattices can be understood
analytically. We have further addressed several
difficulties involved in producing and detecting vortex lattices.

Our
analytical treatment shows that the surface-mode instability is robust
against including relaxation effects in the simple models of pumping
previously considered.  Numerical analysis of the complex Ginsburg-Landau
model reveals that both non-circular geometry and disorder may prevent
formation of a vortex lattice, in contrast to its robustness against
other modifications, such as relaxation, interactions with reservoir
polaritons and higher-order nonlinearities. Sufficient elliptic
deformation or disorder results in a chaotic, nonstationary state. 
However, small deviations from
circular symmetry as well as a small amount of
disorder do allow a regular
vortex lattice to form.  The critical strength of disorder
for $\alpha=4.4, \sigma=0.3$ is
estimated at $\sim 6.5\hbar\omega$, which corresponds to $\sim 0.4$~meV 
at twice-threshold pumping 
in
the stress trap of Ref.~\onlinecite{balili2007:science}.

We have further shown that spontaneously formed, rotating vortex
lattices in a polariton condensate would be possible to detect using
an experimentally feasible defocused homodyne scheme to generate
stable interference images.  By using a defocused homodyne
scheme one can ensure that the same image is generated in each
realization of the vortex lattice.  Furthermore, the images contain
direct information relating to the structure of the vortex lattice,
allowing one in principle to establish the number of vortices and their
arrangement within the lattice.

\begin{acknowledgments}
  MOB acknowledges funding from the Leverhulme Trust.
  JK acknowledges funding from EPSRC grants EP/G004714/2 and
  EP/I031014/1.
  GF acknowledges funding from Marie Curie Actions ESR grants. GF and
  NGB acknowledge funding from EU CLERMONT4 235114. 

\end{acknowledgments}


\appendix

\section{Linear stability analysis}
\label{sec:line-stab-analys}

This appendix provides further details of the analytic stability
analysis presented in Sec.~\ref{sec:inst-non-rotat}.
Sec.~\ref{sec:pert-theory-non} discusses the perturbation theory
for Eq.~(\ref{eq:linearised}), while
Sec.~\ref{sec:exact-zeroth-order} gives the exact zeroth order
solutions of Eq.~(\ref{eq:hypgeom}).  Sec.~\ref{sec:eval-overl-hyperg} 
then shows how the resulting integrals can
be analytically evaluated, and Sec.~\ref{sec:line-stab-gener}
discusses the effects of relaxation terms that may be added to the cGLE.

\subsection{Perturbation theory for Eq.~(\ref{eq:linearised})}
\label{sec:pert-theory-non}

As briefly noted in Sec.~\ref{sec:inst-non-rotat}, the zeroth order
differential operator $i\mathcal{L}$ is not self adjoint, and so
evaluation of perturbation theory requires one to separately determine
the left and right eigenfunctions of this operator.  With such
eigenfunctions identified, one may write the linear part of
Eq.~(\ref{eq:perturbation}) as
\begin{displaymath}
  \delta \omega_i  \psi^R_{0i} 
  + 
  \omega_{0i}   \delta \psi^R_i
  =
  i\mathcal{L} \delta \psi^R_i
  +
  \alpha i \mathcal{\delta{L}}
  \psi^R_{0i}. 
\end{displaymath}
Following standard perturbation theory, one then writes
\begin{multline}
  \label{eq:perturbation-2}
  \delta \omega_i \int d^2r\, \psi^L_{0i} \psi^R_{0i} 
  =
  i \alpha  
  \int d^2r\, \psi^L_{0i}
  \mathcal{\delta{L}}
  \psi^R_{0i} 
  \\+
  \int d^2r\, \psi^L_{0i}
  \left( 
  i\mathcal{L} - \omega_{0i}
  \right)\delta \psi^R_i.  
\end{multline}
The last term will cancel as long as the operator $i \mathcal{L}$ has left
eigenfunctions which satisfy
\begin{displaymath}
  I=
  \int d^2r\, \psi^L_{0i} i\mathcal{L} \psi
  =
  \omega_{0i}
  \int d^2r\, \psi^L_{0i}  \psi
\end{displaymath}
for all $\psi$. 

The right eigenfunctions are given directly by solutions of
Eq.~(\ref{eq:linearised}). To find the corresponding left
eigenfunctions, it is convenient to write out the components of
the eigenfunctions as $\psi^R_{0i} = (h^R_{0i}, u^R_{r,0i},
u^R_{\theta_{0i}})^T$ and $\psi^L_{0i} = (h^L_{0i}, u^L_{r,0i},
u^L_{\theta_{0i}})$ where $u_r, u_\theta$ are radial and polar
components of the velocity vector $\vec{u}$.  The zeroth order right
eigenfunctions obey Eq.~(\ref{eq:linearised}) with $\vec{v}=0,
\alpha=0, \sigma=0$.  This can be written out in components as
\begin{equation}
  \label{eq:right-evecs}
  \begin{aligned}   
     \omega_{0i} h^R_{0i} 
    &=
    -
    \frac{i}{r} \frac{d}{dr} (r \rho u^R_{r,0i})
    + 
    \frac{ s}{r} \rho u^R_{\theta,0i},
    \\
    \omega_{0i} u^R_{r,0i} &= - \frac{i}{2} \frac{d h^R_{0i} }{dr},
    \qquad
    \omega_{0i} u^R_{\theta,0i} = \frac{ s}{2r} h^R_{0i}.
  \end{aligned}
\end{equation}
Writing out the components $\psi = (h,u_r,u_\theta)^T$, one has
\begin{multline*}
  I\!=\!\!
  \int \!d^2r 
  \left[
    -
    h^L_{0i}
    \frac{i}{r} \frac{d}{dr} (r \rho u_{r})
    + 
    h^L_{0i}
    \frac{ s}{r} \rho u_{\theta}
  \right.\\\left.
    -
    u^L_{r,0i}
    \frac{i}{2} \frac{d h }{dr} 
    +
    u^L_{\theta,0i}
    \frac{ s}{2r} h
  \right].
\end{multline*}
Integrating by parts (noting that the integral range is
$0<r<\sqrt{\mu}$ and that $r\rho(r)$ vanishes at the boundary), one
may show that if $(h^L_{0i}, u^L_{r,0i}, u^L_{\theta_{0i}}) =
(h^{R\ast}_{0i}, 2 \rho u^{R\ast}_{r,0i}, 2 \rho
u^{R\ast}_{\theta_{0i}})$ then
\begin{displaymath}
  I = \omega_{0i} \int d^2r
  \left[
    h^{R\ast}_{0i} h
    +
    2 \rho \left(
      u^{R\ast}_{r,0i} u_r
      +
u^{R\ast}_{\theta,0i} u_{\theta}
\right)
\right]
\end{displaymath}
as required. Hence the above gives the relation between
the left and right eigenfunctions.

\subsection{Exact zeroth order solutions}
\label{sec:exact-zeroth-order}

With the left eigenfunctions as defined above, the shift in frequency
can be explicitly determined by evaluating
Eq.~(\ref{eq:perturbation-2}).  The functions $u^R_{r,0i},
u^R_{\theta_{0i}}$ are given in terms of $h^R_{0i}$ in
Eq.~(\ref{eq:right-evecs}), and substituting into the equation for
$h^R_{0i}$ yields Eq.~(\ref{eq:hypgeom}).  For $\rho = \rho_{\rm TF}$, and
$h^R_{0i}(r,\theta) = e^{is\theta} h^R_{0i}(r)$, this equation is
hypergeometric.  By restricting to solutions that remain bounded as
$r^2 \to \mu$, one finds the solutions to be the finite order
hypergeometric polynomials, with frequencies as given in
Eq.~(\ref{eq:eigenfrequencies}) and wavefunctions
\begin{equation}
  h^R_{0i}(r,\theta)
  =
   e^{i s\theta} r^s
  {}_2F_1\left(-n,s+1; n+s+1; r^2\right).
\end{equation}
This simple form is the two-dimensional analogue of the results given
in three dimensions in Refs.~\onlinecite{Stringari1996} and
\onlinecite{Stringari1998}.  Note that 
the two-dimensional harmonic trap is \emph{not} equivalent to the
highly anisotropic pancake limit of the three-dimensional trap---the
two dimensional case presented here has a hard wall boundary condition
in the $z$ direction.  As a result, the excitation spectrum given here
is not the same as in Ref.~\onlinecite{Stringari1998}.

\subsection{Evaluating overlaps for hypergeometric functions}
\label{sec:eval-overl-hyperg}

Using the results above, one finds that
\begin{multline}
  \label{eq:integral-overlaps}
    \int d^2r\,
    \psi^L_{0i} \delta\mathcal{L} \psi^R_{0i}
    \\=
    \int d^2r
    \left[\vphantom{\frac{dh}{dr}}
     (\alpha - 2\sigma \rho) |h^R_{0i}|^2 
    - \frac{ h^{R\ast}_{0i}}{r} \frac{d}{dr} (r h^R_{0i} v) 
   \right. \\ \left.
    - \frac{ \rho}{2 \omega^2} \frac{dh^{R\ast}_{0i}}{dr} \frac{d}{dr} 
    \left( v \frac{dh^R_{0i}}{dr} \right)
    -
    \frac{ s^2 v \rho h^{R\ast}_{0i}}{2 \omega^2 r^2}   \frac{dh^R_{0i}}{dr}
  \right].
\end{multline}
By noting that $v(r) = |\vec{v}(r)|$ vanishes at the limits, one may
integrate the second and third terms by parts.  Using 
Eq.~(\ref{eq:hypgeom}), one then finds that Eq.~(\ref{eq:integral-overlaps}) 
simplifies to
\begin{equation}
  \label{eq:final-shift-integral}
  \int d^2r\,
  \psi^L_{0i} \delta\mathcal{L} \psi^R_{0i}
  =
  \int d^2r\,
  [\alpha - 2\sigma (\mu-r^2)] |h^R_{0i}|^2.
\end{equation}
 Hence, making use of $\mu = 3 \alpha/2\sigma$,
 Eq.~(\ref{eq:change}) becomes
\begin{math}
  \delta \omega_i = 
  (i \alpha/2) \left( 3 \langle r^2 \rangle_i/\mu - 2 \right),
\end{math}
where
\begin{displaymath}
\langle r^2 \rangle_i = 
\left. \int d^2r\, r^2 |h^R_{0i}|^2  \middle/ \int d^2r\,  |h^R_{0i}|^2. \right.
\end{displaymath}
This integral can be found straightforwardly by using the identity
\begin{displaymath}
    0 =
    \int\limits_0^{\sqrt{\mu}} \!\! dr \frac{d}{dr}
  \left[ \left( \rho r \frac{dh}{dr} \right)^2 \right] 
  =
  \int\limits_0^{\sqrt{\mu}} \!\! dr
  \frac{d}{dr}
  \left[
    \frac{s^2 \rho^2}{2}
    - \omega^2 \rho r^2
  \right] h^2
\end{displaymath}
(making use of Eq.~(\ref{eq:hypgeom})). Evaluating the second term, and
using the eigenvalues equation, one finds that
\begin{math}
  \langle r^2 \rangle_{i} = \mu (s^2 + \omega_i^2) / (s^2 + 2 \omega_i^2).
\end{math}
Using Eq.~(\ref{eq:eigenfrequencies}) for the eigenfrequencies, one
recovers the result in Eq.~(\ref{eq:im-shift}).

\subsection{Stability of the generalized order parameter equation}
\label{sec:line-stab-gener}

Including the phenomenological relaxation term discussed in
Sec.~\ref{sec:incl-phen-relax} the cGLE becomes
\begin{equation}
  \label{eq:cgpe2}
  \begin{split}
    i \frac{\partial\psi}{\partial t} = \frac{1}{2} \Big[ &- \nabla ^2
    +
    V(x,y) +| \psi | ^2\\
    &+ i \Big( \alpha (x,y) - \sigma |\psi| ^2 - 2 i \eta \partial_t
    \Big) \Big ] \psi,
  \end{split}
\end{equation} 
and the Madelung representation, Eq.~(\ref{eq:madellung}), becomes
\begin{equation}
  \label{eq:madellung2}
  \begin{split}
  &\partial_t \rho + \nabla\cdot(\rho \vec{v})
  = (\alpha - \sigma \rho + 2 \eta \partial_t \phi) \rho
  \\
  &2 \partial_t \phi +  ( \rho + r^2 + |\vec{v}|^2) 
  = 0.
  \end{split}
\end{equation}
The second of these equations has now been written explicitly for the
phase, rather than for $\vec{v} = \nabla \phi$, since the phase now
appears explicitly in the first equation.  In keeping with the
hydrodynamic approximation which neglects $\nabla^2 \sqrt{\rho} /
\sqrt{\rho}$, a similar term $\eta \partial_t \rho/\rho$ has also been
neglected.

Since the second of Eq.~(\ref{eq:madellung2}) is unchanged, the
stationary density profile remains as before ($\vec{v}$ is of first
order in pumping and decay terms), with $\mu = -2 \partial_t \phi$.
The continuity equation is however modified, so that integrating it over the entire space gives
 $\int 2 \pi r
dr [\alpha - \sigma (\mu-r^2) - \eta \mu] (\mu-r^2) = 0$, yielding
$\mu = 3 \alpha/(2 \sigma + 3 \eta)$.

The linearized fluctuation equations are complicated by the explicit
appearance of $\phi$, so it is necessary to expand $\phi \to \phi
+ \varphi(t)$; one finds
\begin{multline}
  \label{eq:relaxation-perturbation}
  \left(
    \begin{array}{cc}
      1 & - 2 \eta \rho \\ 0 & 1
    \end{array}
  \right)
  \partial_t \left(
    \begin{array}{c}
      h \\ \varphi
    \end{array}
  \right)
  =
  \left(
    \begin{array}{c}
      - \nabla\cdot (\rho \vec{u} )
      \\
      - h/2
    \end{array}
  \right)
  \\
  +
  \left(
    \begin{array}{c}
      - \nabla \cdot (h \vec{v})
      +
      \left( \alpha - 2\sigma \rho - \eta \mu\right) h  
      \\
       - \vec{v} \cdot \vec{u}
    \end{array}
  \right),
\end{multline}
where $\vec{u} = \nabla \varphi$.  In the above, the time derivatives
have been kept together on the left hand side.  One may then proceed
by multiplying both sides by the inverse of the matrix appearing on
the left, and expanding to first order in the pumping and decay terms
$\alpha$, $\sigma$, and $\eta$.  The net result of this procedure is that the
expression $(\alpha - 2 \sigma \rho)$, which appeared in
Eq.~(\ref{eq:final-shift-integral}), is now replaced by $(\alpha - 2
\sigma \rho) \to (\alpha - 2 \sigma \rho - \eta \mu - \eta \rho)$.
Hence the frequency shift is given by
\begin{math}
  \delta \omega_i = (i/2) \left[
    \alpha - 2(\sigma + \eta) \mu + (2 \sigma + \eta) \langle r^2 \rangle_i
  \right].
\end{math}
Combining this with the values of $\mu$ and $\langle r^2 \rangle_i/\mu$ given
above gives Eq.~(\ref{eq:im-shift-eta}).

\section{Disorder potential}
\label{sec:Vdis}
This appendix gives technical details for the inclusion of disorder
discussed in Sec.~\ref{sec:disorder}. The disorder is included in the
external potential by letting \mbox{$V(\vec{r}) \rightarrow
V(\vec{r})+V_{\rm dis}(\vec{r})$}, where $V_{\rm dis}(\vec{r})$ is a
Gaussian random field, constructed on the computational grid as a
Fourier sum $V_{\mathrm{dis}}=\sum_{i,j}(a_{ij}/2)\exp\left(
i\left(k_i x+k_j y+\phi_{ij}\right)\right)$, whose amplitudes 
$a_{ij}=a_{-i-j}$ are sampled from a Gaussian distribution with mean
zero and standard deviation
\begin{equation}
  \label{std} 
  s_{ij}=v_0\exp\left[-l_c^2\left(k_i^2+k_j^2\right)\right]. 
\end{equation} 
The phase  $\phi_{ij}$ is a random number in the interval $[0,2\pi]$.
The quantities $v_0$ and $l_c$, defined for computational convenience,
are related to the physical disorder strength $V_0$ and correlation
length $\xi$ by noting that
\begin{equation}
  \label{eq:real-space-corr}
  \langle V_{\mathrm{dis}}(\vec{r})V_{\mathrm{dis}}(\vec{r}^{\,\prime})\rangle 
  \simeq 
  \frac{v_0^2}{4(\Delta k)^2}\frac{\pi}{2l_c^2}
  \exp\left[-\frac{\left(\vec{r} - \vec{r}^{\,\prime} \right)^2}
    {(8l_c^2)}\right],
\end{equation}
and identifying $\xi=2l_c$ and $V_0=v_0\sqrt{\pi/2}/(2l_c\Delta k)$,
where $\Delta k$ is the numerical Fourier mode
spacing. Eq.~(\ref{eq:real-space-corr}) follows from the fact that  
$\langle a_{ij}a_{uv} e^{i\left(\phi_{ij}+\phi_{uv}\right)}\rangle
= \delta_{iu}\delta_{jv}\langle a_{ij}^2 e^{2\phi_{ij}}\rangle 
+ \delta_{i-u}\delta_{j-v}\langle a _{ij}^2 \rangle 
= \delta_{i-u}\delta_{j-v}s^2_{ij}$, and replacing $k$-space sums
by integrals.
  
\section{Constructing interference images}
\label{sec:interference-images}

This appendix gives additional technical detail on the calculation of
interference images. Vortex-lattice solutions are generated by
propagating Eq.~(\ref{eq:cgpe}). 
The general principle for interference imaging of rotating vortex 
lattices discussed in Sec.~\ref{sec:detect-rotat-vort} does not, however, depend
on the specific Ginzburg-Landau model.  The dispersion spectrum of the
condensate is 
found as the modulus squared of the 3D Fourier transform of the
condensate wavefunction sampled over a period of time:
\begin{equation}
  \label{eq:spectrum}
  S(\omega,\mathbf{k}) =
  \left.\left|\int d^2r\,e^{-i\mathbf{k}\cdot\mathbf{r}}
  \int dt\,e^{-i\omega t}\psi(\mathbf{r},t)\right|\right.^2.
\end{equation}

The time-integrated interference images are found as the integral
\begin{equation}
  \label{eq:interference}
  I(\vec{r}) = \int_T dt\,
  \left|\psi(\vec{r}+\vec{r}_\mathrm{d},t) + 
  \psi_{\rm ref}(\vec{r}-\vec{r}_\mathrm{d},t)\right|^2,
\end{equation}
where the time $T$ is the time over which light is gathered in the
measurement. Here we choose $T$ to correspond to a few
revolutions of the vortex lattice.  Since the interference image is
stable, it is insensitive to the choice of $T$. The form of
$\psi_{\rm ref}$ is given by the choice of imaging method
[Secs.~\ref{sec:heterodyne}--\ref{sec:defocused}]. The displacement
$\vec{r}_\mathrm{d}$ is non-zero only for the homodyne
scheme in Sec.~\ref{sec:homodyne}. 

For the frequency-filtered homodyne and, optionally, the
defocused homodyne 
schemes, the light gathered from the condensate is assumed to be
passed through a band-pass filter that suppresses Fourier components
outside the interval $[\omega_{\rm low},\omega_{\rm high}]$, creating 
\begin{equation}
  \label{eq:filter}
  \begin{split}
    \psi_{\rm filt}&(\vec{r},t) \\
      &=\frac{1}{2\pi}
      \int_{\omega_{\rm low}}^{\omega_{\rm high}}d\omega\,\left[ 
      e^{i\omega t}
      \int_{-\infty}^{\infty} dt\, e^{-i\omega t}\psi(\vec{r},t)\right],
  \end{split}
\end{equation}
which is then used to create the interference
  image~(\ref{eq:interference}).


%

\end{document}